\documentclass[11pt]{article}
\usepackage{amssymb}
\usepackage{amsmath} 
\usepackage{amsthm}
\usepackage{bm}
\usepackage{subcaption}
\usepackage{graphicx}

\begin{document}

\begin{titlepage}
\begin{center}
\textbf{Theory and investigation of acoustic multiple-input multiple-output systems based on spherical arrays in a room} \\ \vspace{10ex}
Hai Morgenstern$^{1,}$\footnote{e-mail: haimorg@ee.bgu.ac.il}, Boaz Rafaely$^1$, and Franz Zotter$^2$\\
$^{1}$Department of Electrical and Computer Engineering, \\
Ben Gurion University of the Negev, Beer- Sheva, 84105, Israel\\
$^{2}$ Institute of Electronic Music and Acoustics \\
University of Music and Performing Arts, \\
Graz, Austria\\
\end{center}
\end{titlepage}
		
\begin{abstract}
The spatial attributes of a sound field affect the perception of the sound by human listeners, motivating further study of spatial attributes of room acoustics, beyond the one provided in ISO 3382.  
Microphone and loudspeaker arrays have been separately studied for this purpose, though, systems that combine both, referred to as multiple-input multiple-output (MIMO) systems, can potentially provide a more powerful tool for room acoustics analysis due to the ability to simultaneously control both arrays. 
Nevertheless, MIMO systems have been studied to a limited degree in the context of room acoustics. 
This paper offers a theoretical framework for the spatial analysis of enclosed sound fields using a MIMO system comprising spherical loudspeaker and microphone arrays.  
A system transfer matrix is formulated for free-field conditions and then generalized for rooms using an image source method.  
System properties are studied; the system singular vectors are related to the directions of arrival and radiation at the microphone and loudspeaker arrays, respectively, and the rank of the system is related to the number of significant reflections. 
Simulation and experimental studies support the theory and validate the spatial properties of acoustic MIMO systems in rooms. 
				
\end{abstract}

\section{INTRODUCTION}

According to ISO 3382, the reverberation time of a room is recognized as the most important measure for the acoustics of the room \cite{standard19973382}. 
The ISO standard also presents other measures, such as early-to-late energy ratios and clarity, providing a more complete evaluation of the acoustics of a room, yet this evaluation is in terms of energy and is lacking in spatial information. 
Spatial attributes of the sound field in a room affect the perception of the sound by a human listener \cite{marshall2001spatial,griesinger1998general,morimoto2001role}, implying that valuable information regarding the acoustics of a room cannot be extracted using only energetic measures. 
In this paper, we aim to expand the suite of available methods for the acoustical evaluation of rooms to include measurement configurations comprised of spherical loudspeaker and microphone arrays that enable a spatial analysis around both arrays.  


Spatial measures such as the directivity index, directional diffusion and directional decay rates \cite{gover2002microphone} use spatial sound field information to better predict the acoustical performance of enclosures, facilitating the design and analysis of enclosures with respect to the perception of speech and music. 
Computing spatial measures in an auditorium requires measurement systems with directional microphones or microphone arrays, typically placed in the seating area. 
Microphone arrays have been extensively studied for room acoustics analysis, for detecting the direction of arrival (DOA) of early room reflections and for computing directional impulse responses \cite{rafaely2005analysis, gover2002microphone}. 
In particular, spherical microphone arrays have been proposed for studying room acoustics due to the complete 3-D analysis they facilitate \cite{balmages2004room}. 
Microphone arrays, however, provide spatial analysis with somewhat limited resolution; 
for example, in the study of directional impulse responses captured with a microphone array, spatially dense reflections may be identified as a single reflection source if the reflections have directions of arrival (DOAs) with an angle difference smaller that the array's spatial resolution. 
These measurements are typically conducted using an omnidirectional loudspeaker. 
An omnidirectional loudspeaker radiates sound in all directions equally, therefore exciting all reflection paths from the loudspeaker to the microphones, which can potentially yield a dense reflection pattern at the microphone array. Thus, the use of microphone arrays and an omnidirectional loudspeaker limits the spatial resolution in the analysis of a room impulse response (RIR).


Directional loudspeakers have been shown to have advantages over omnidirectional ones in room acoustic measurements due to their ability to control the excitation of individual reflection paths and to improve the separability of individual reflections \cite{tervo2009acoustic}.  
In particular, spherical loudspeaker arrays, comprised of a matched set of loudspeaker units mounted on the surface of a sphere, are good candidates for realizing directional loudspeakers. 
They can be modeled as compact sources with adjustable directivity that can be steered over all directions \cite{rafaely2009spherical}, capable of producing complex radiation patterns in spite of the fact that they are physically compact. 
As stated earlier, typical measurement systems for room acoustics analysis, such as found in refs. 
[5], [10], [11] and [7], provide a limited spatial analysis since these systems are comprised of either a single loudspeaker and a microphone array or a single microphone and a loudspeaker array. 


In using systems that combine microphone and loudspeaker arrays, the benefits of each are retained, and the ability of simultaneously controlling both arrays provides additional degrees of freedom.    
This idea originated in Gerzon's famous contribution \cite{gerzon1975recording}. 
In 1975, Gerzon proposed recording a set of RIRs between several loudspeakers on a stage and a spherical microphone array \cite{gerzon1975recording}.  
Farina expanded Gerzon's proposal and presented a novel measurement method for capturing spatial characteristics using cardioid microphones and a sound field microphone mounted on a rotating boom \cite{farina2003recording}. 
In a follow up paper, another measurement technique was presented in which a set of RIRs was recorded using spherical arrays of loudspeakers and microphones \cite{farina2006room}. 
Farina's approach removed the limitations of Gerzon's approach - omnidirectional sources were generalized to directional sources, realized using spherical loudspeaker arrays, and microphone arrays with higher spatial resolution were used.  
Farina's work focuses on the method of recording these spatial RIRs, suggesting that a collection of the proposed MIMO measurements may be able to deepen the understanding of the spatial properties of sound fields. 
However, no theoretical investigation of the MIMO system and dedicated analysis methods were presented.


Some theory and algorithms regarding MIMO systems based on spherical arrays have been published by the authors in conferences; 
[15] 
presents a limited formulation of an acoustic MIMO system, which will be extended in this contribution.
Moreover, [16] and [17] 
present methods for finding directions of radiation and arrival at the loudspeaker and microphone arrays, respectively. 
These papers show the potential of using MIMO systems as a platform for studying room acoustics.


This paper focuses on the development of a theoretical framework for the spatial analysis of enclosed sound fields, using an acoustic MIMO system comprised of a spherical loudspeaker array and a spherical microphone array. 
The main contributions of the paper are: 
\begin{itemize}
\item A theoretical formulation of an acoustic MIMO system for free-field conditions. 
An analytical, closed-form description of the responses between the microphone array and the loudspeaker array is formulated for a variety of microphone array configurations, such as open sphere and rigid sphere, based on a simplifying far-field assumption\cite{morgenstern2014farfield}. 
\item An analytic study of the system's spatial properties.  
The closed-form description of the far-field model enables an analytic study of the system's spatial properties. 
The system is shown to have unit rank, and the system's singular values are shown to be invariant to rotation and mirroring of both loudspeaker and microphone arrays, demonstrating the system's spherical isotropy. 
\item Generalization to an acoustic system model in enclosures.
Following the image method\cite{allen1979image}, the room reflections are replaced by loudspeaker array image sources. 
The rank and the singular vectors of the system model for this case are studied. 
\item A comprehensive simulation and experimental study, validating the spatial properties of an acoustic MIMO system for simulated rooms with adjustable absorption coefficients and for a real lecture hall. 
\end{itemize}

After a brief background on the spherical Fourier transform (SFT) in section \ref{sec:background_SH}II, acoustic MIMO systems under free-field conditions and for rooms are formulated and studied in sections \ref{sec:sysmodelFF}III-\ref{sec:FF2RTF}V.
The paper concludes with a simulation study and an experimental investigation in sections \ref{sec:SimStud}VI and \ref{sec:EXPInvestigation}VII.  

\section{SPHERICAL HARMONIC DECOMPOSITION}
\label{sec:background_SH} 

Spherical harmonic (SH) decomposition is employed in this paper to represent sound fields, both radiated by the loudspeaker array and measured at the microphone array. 
Therefore, the spherical Fourier transform (SFT) \cite{driscoll1994computing}, or SH decomposition \cite{williams1999fourier}, is now briefly introduced.


The standard spherical coordinate system \cite{arfken1999mathematical}, $(r,\theta,\phi)$, is used throughout the paper. 
A function, $f(\theta,\phi)$, which is square integrable on the sphere, can be represented as a linear combination of SH functions using the SFT of the function, $f_{nm}$, given by\cite{driscoll1994computing}:
\begin{eqnarray}\label{eq:back_SFT}
f(\theta,\phi) & = & \sum_{n=0}^{\infty}\sum_{m=-n}^{n} f_{nm}Y_n^{m}(\theta, \phi),
\end{eqnarray} 
in which $Y_n^{m}(\cdot,\cdot)$ are the SH functions, defined by:
\begin{eqnarray}\label{eq:sh_def}
Y_n^{m}(\theta, \phi) & = &  \sqrt{ \frac{(2n+1)(n-m)!}{4\pi (n+m)!}}P_n^{m}(\cos\theta)e^{jm\phi},
\end{eqnarray}
with $n$, $m$ being the SH order and degree, respectively, $P_n^{m}(\cdot)$ is the associated Legendre function, and $j=\sqrt{-1}$. 
For convenience, boldface angles are vector notation for elevation and azimuth angles; 
$Y_{n}^{m}(\bm \theta)$ stands for the SH function evaluated at $\bm \theta = (\theta,\phi)$.  

When a set of samples of a function is given, the SFT can be computed by a weighted summation over the samples for order-limited functions, i.e., functions with $f_{nm} = 0, \, \forall n>N$, with $N$ being the SH order of the function.
In the special case of a nearly-uniform distribution\cite{rafaely2005analysis} of $L$ samples at sampling positions $\bm \theta_j$, the SFT is computed using a regular summation:
\begin{eqnarray}\label{eq:DiscreteSFT1}
f_{nm} & = &  \frac{4\pi}{L} \sum_{l=1}^{L} \; f(\bm \theta_l)Y_n^{m^*}(\bm \theta_l).
\end{eqnarray}  
The reader is referred to refs. [6] and [23] 
for further reading on sampling functions on a sphere.

To formulate the relations from eqs.~\eqref{eq:back_SFT} and \eqref{eq:DiscreteSFT1} in matrix form, two vectors are defined: 
\begin{eqnarray}\label{eq:DiscreteSFT2}
\bm f & = & [ f(\bm \theta_1), \, f(\bm \theta_2), \, ... \, f(\bm \theta_L)  ]^\mathrm{T}, \\ \label{eq:DiscreteSFT3}
\bm f_{SH} & = & [ f_{00}, \, f_{1-1}, \, f_{10}, \, f_{11}, \, ... \, f_{NN}  ]^\mathrm{T},
\end{eqnarray} 
which are a vector with the values of the function at the sampling positions and a vector of the $(N+1)^2$ SH coefficients of its SFT, respectively. 
Eq.~\eqref{eq:back_SFT}, calculated at the sampling positions, and eq.~\eqref{eq:DiscreteSFT1} can now be written as:
\begin{eqnarray}\label{eq:DiscreteSFT4}
\bm f & = & \bm Y_{L} \bm f_{SH}, \\
\bm f_{SH} & = &\frac{4\pi}{L} \bm Y_L^\mathrm{H} \bm f, \label{eq:DiscreteSFT5}
\end{eqnarray} 
where $\bm Y_{L} = [ \bm y_1, \, \bm y_2, \, ... \, \bm y_L ]^\mathrm{T}$ is a $L\times (N+1)^2$ matrix, and  
\begin{eqnarray}\label{eq:DiscreteSFT4a}
\bm y_l = [Y_0^{0}(\bm \theta_l), \, Y_1^{-1}(\bm \theta_l), \, ... \, Y_N^{N}(\bm \theta_l)]^\mathrm{T}
\end{eqnarray} 
is an $(N+1)^2\times 1$ vector. 
The latter formulation also suggests that the singular values of $\sqrt{\frac{4\pi}{L}  } \bm Y_L$ all have a value of unity when considering nearly-uniform distributions.   
Eqs.~\eqref{eq:DiscreteSFT5} and \eqref{eq:DiscreteSFT4} are regarded as discrete versions of the SFT and its inverse, respectively. 

Lastly, the following SFT identity will be used later: 
\begin{eqnarray}\label{eq:SFTIDENTITY}
	 \int_{0}^{2\pi}\int_{0}^{\pi}    Y_n^m(\bm \theta)\delta(\bm \theta - \bm \theta_i)\sin \theta d\theta d\phi &=&
	 Y_n^m(\bm \theta_i), 
\end{eqnarray}
where $\delta(\bm \theta - \bm \theta_i) = \delta(\cos\theta- \cos\theta_i)\delta(\phi- \phi_i)$ is the Dirac delta function.   


\section{MIMO SYSTEM MODEL IN FREE-FIELD}\label{sec:sysmodelFF}

An acoustic MIMO system is comprised of a spherical loudspeaker array with $L$ loudspeakers distributed over a sphere with a radius of $r_L$ and a spherical microphone array with $M$ microphones distributed over a sphere with a radius of $r_M$. 
To calculate the sound field radiated by an actual spherical loudspeaker array, the array is modeled by a rigid sphere covered by $L$ spherical caps, each imposing a radial velocity of $u_l$ for $l = 1, 2, ..., L$. 
This model, which is employed due to the simple analysis it facilitates, has been previously validated\cite{meyer2000multi, zotter2007modeling}.
Assuming this model, the sound field measured at a distance $r$ and spatial angles $\bm \theta$ with respect to the loudspeaker array center is given by\cite{williams1999fourier}: 
\begin{eqnarray}\label{eq:sys1}
p(k,r,\bm \theta) & = & j\rho_0 c \sum_{n=0}^{\infty} q(n,\cos\alpha) \frac{h_n(kr)}{h_n'(kr_L)}   \nonumber \\ 
&  & \times \sum_{m=-n}^{n} \, Y_n^m(\bm \theta) \sum_{l=1}^{L}u_l\, Y_n^{m^*}(\bm \theta_l)  , 
\end{eqnarray}
where $k$ is the wavenumber, $\rho_0$ is the air density coefficient, $c$ is the speed of sound, $h_n(\cdot)$ and $h'_n(\cdot)$ are the spherical Hankel function of the first kind with order $n$ and its derivative, respectively, $r_L$ is the radius of the loudspeaker array, $P_{n}(\cdot)$ is the Legendre polynomial, and $u_l$ is the radial velocity of the $l^{th}$ cap,  $\bm \theta_l$ is the vector form for the spatial angles of the $l^{th}$ cap for $l=1,..,L$, and $q(n,\cos\alpha)$ is given by: 
\begin{eqnarray}\label{eq:sys1d}
q(n,\cos\alpha) = \begin{cases}
4 \pi L \left( 1- \cos\alpha \right) & \mbox{, } n=0 \\ \frac{4\pi L }{2n+1} \left(  P_{n-1}(\cos\alpha) -P_{n+1}(\cos\alpha)  \right) & \mbox{, } n>0
\end{cases}.
\end{eqnarray}

Due to the nearly-uniform distribution of the loudspeaker units and following eq.~\eqref{eq:DiscreteSFT5}, the summation over $L$ from eq.~\eqref{eq:sys1} is identified as a discrete SFT of the array's continuous radial velocity: 
\begin{eqnarray}\label{eq:sys1b}
u_{nm}& = &  \frac{4\pi}{L}\sum_{l=1}^{L}u_l\, Y_n^{m^*}(\bm \theta_l). 
\end{eqnarray}  
With $L$ loudspeakers comprising the array, only $L$ SHs in $u_{nm}$ can be controlled, defined over the range of $0\leq n\leq N_L, \; -n\leq m \leq n$, with $N_L$ satisfying $(N_L+1)^2\leq L$. $N_L$ is referred to as the SH order of the loudspeaker array.
 
The radiated sound field is measured by a spherical microphone array at a distant location, nonconcentric to the loudspeaker array. 
In the case of open microphone array configurations using sound pressure transducers, such as the single and dual open-sphere configurations, the microphone signals can be simulated by evaluating eq.~\eqref{eq:sys1} at the individual microphone positions relative to the loudspeaker.
This can be done since the microphone array does not impose any boundary conditions on the radiated field. 
On the other hand, when microphone arrays configured around a rigid sphere are considered, an additional component is added to the sound field, the scattered field.  
In this case, a closed-form description of the sound field is more difficult and might require translation operators or multiple scattering\cite{gumerov2005fast}, a complication which is avoided here by well-justified simplifications. 

A simplified far-field model that facilitates a closed-form analytical description of the system transfer function is more suitable for the analysis of the system properties, and is developed in this paper. 
A criterion for a planar wavefront approximation at a microphone array due to directional sources, has been presented recently\cite{morgenstern2014farfield}. 
Using this approximation, and assuming an incident field composed of a single plane-wave, the plane-wave amplitude, $\tilde{a}$, is derived by calculating the pressure, radiated by the loudspeaker array, at $(D, \bm \theta_{LM})$. 
Following eq.~\eqref{eq:sys1}, this is given by:
\begin{eqnarray}\label{eq:sys1aa}
\tilde{a} & = &  j\rho_0 c \sum_{n=0}^{N_L} q(n,\cos\alpha) \frac{h_n(kD)}{h_n'(kr_L)}  \sum_{m=-n}^{n} \, Y_n^m(\bm \theta_{LM})\, u_{nm}.  
\end{eqnarray} 
Now, a spherical microphone array is introduced, centered at $(D, \bm \theta_{LM})$. 
Given the amplitude $\tilde{a}$, the pressure around the microphone array's center can be expanded in SH by\cite{williams1999fourier}: 
\begin{eqnarray}\label{eq:sys2PWD}
p(k,r_M, \bm \eta) & = &  \sum_{n'=0}^{\infty} \sum_{m'=-n'}^{n'} \tilde{a}\, 
b_{n'}(kr_M) Y_{n'}^{m'^*}(\bm \eta_{ML})Y_{n'}^{m'}(\bm \eta), 
\end{eqnarray}
in which $r_M$ is the radius of the microphone array, $\bm \eta$ are the elevation and azimuth angles with respect to the microphone array center, $\bm \eta_{ML}$ is the DOA of the plane-wave, i.e., the direction of the loudspeaker array center with respect to the microphone array center, and $b_{n}(\cdot)$ is generalized for open and rigid spheres as follows\cite{rafaely2004plane}: 
\begin{eqnarray}
b_n(kr) & =& \begin{cases}
4\pi j^n\left[  j_n(kr)- \frac{j'_n(kr_0)}{h'_n(kr_0)} h_n(kr)  \right] & \text{rigid sphere,} \\ 
4\pi j^n j_n(kr) & \text{open sphere,}
\end{cases}
\end{eqnarray} 
where $j_{n}(\cdot)$ and $j'_{n}(\cdot)$ are the spherical Bessel function of order $n$ and its derivative and $r_0\leq r_M$ is the radius of the rigid sphere.
The analytic generalization to different microphone array configurations is one of the advantages of the far-field simplified model. 

The pressure at the microphones is calculated using eqs.~\eqref{eq:sys1aa} and \eqref{eq:sys2PWD} by substituting their positions with regard to the array's center, $(r_{M}, \bm \eta_{i})$, given by:
\begin{eqnarray}\label{eq:sys3PWD}
p(k,r_{M}, \bm \eta_{i}) & = &  j\rho_0 c
\sum_{n=0}^{N_L} \sum_{m=-n}^{n} \,  \frac{h_n(kD)  q(n,\cos\alpha) }{h_n(kr_L)} u_{nm} Y_n^m(\bm \theta_{LM}) \times \nonumber  \\
& & 
\sum_{n'=0}^{\infty}\sum_{m'=-n'}^{n'} \, b_{n'}(kr_M)  Y_{n'}^{m'^*}(\bm \eta_{ML}) Y_{n'}^{m'}(\bm \eta_{i}). 
\end{eqnarray}

Eq.~\eqref{eq:sys3PWD} combined with eq.~\eqref{eq:sys1b} describes the transfer functions between the $L$ loudspeaker units and the $M$ microphones. 
To formulate the system model in matrix form, first, the infinite summation over $n'$ in eq.~\eqref{eq:sys3PWD} is limited to the achievable order of the microphone array, $N_M$, by considering practical limitations such as the number of microphones and the array's radius\cite{rafaely2005analysis}. 
Then, two vectors are defined, an $M\times 1 $ vector of the sampled pressure at the microphones and an $L\times 1 $ vector of the velocity of the loudspeaker units:
\begin{eqnarray}\label{eq:sys1di}
\bm p(k) & = & [   p_1, \, p_2,\,  ... \, p_M    ]^\mathrm{T}, \\ \label{eq:sys1dii}
\bm u(k) & = & [ u_1, \, u_2,\, ... \, u_L    ]^\mathrm{T}.
\end{eqnarray}
Following eq.~\eqref{eq:sys3PWD}, the system is presented in matrix form by:
\begin{eqnarray}\label{eq:sysclassic}
\bm p(k) &=& \bm G_s(k) \bm u(k),
\end{eqnarray}
with: 
\begin{eqnarray}\label{eq:sysPWDM1}
\bm G_s(k) & = & \bm Y_{M} \bm B_M \bm y^{*}(\bm \eta_{ML})\bm y^\mathrm{H}(\bm \theta_{LM}) \bm H_L(D) \bm Q_L \frac{4\pi}{L} \bm Y_L^\mathrm{H}, 
\end{eqnarray}  
where $\bm y(\bm \theta_{LM})$ and $\bm y(\bm \eta_{ML})$ are the steering vectors of the loudspeaker and microphone arrays, as given in eq.~\eqref{eq:DiscreteSFT4a}, with relative directions at loudspeaker and microphone arrays, $\theta_{LM}$ and $\eta_{ML}$, having dimensions of $(N_L+1)^2\times 1 $ and $(N_M+1)^2\times 1 $, respectively;
$\bm Y_{L}$ and $\bm Y_{M}$ are the steering matrices of the loudspeaker and microphone arrays, with the directions of the individual loudspeakers and the microphones with respect to the loudspeaker and microphone array center as given in eq.~\eqref{eq:DiscreteSFT4a}; 
$\bm B_M$, $\bm H_L(D)$, and $\bm Q_L$ are diagonal matrices with dimensions of $(N_M+1)^2\times (N_M+1)^2$, $(N_L+1)^2\times (N_L+1)^2$, and $(N_L+1)^2\times (N_L+1)^2$, respectively. 
These matrices are defined by their diagonals, given by:
\begin{eqnarray}\label{eq:sysFFMIMO1}
\bm B_M & = &  \mbox{diag}[\bm b_M],\,\,\\
\mbox{with } \bm b_M & = & [b_0(kr_M),\, b_1(kr_M),\,  b_1(kr_M),\,  b_1(kr_M),...,\,  b_{N_M}(kr_M)], \nonumber
\end{eqnarray}
\begin{eqnarray}\label{eq:sysFFMIMO2}
\bm H_L(D) & = &  j\rho_0 c\times \,\, \mbox{diag}[\bm h_L(D)], \,\, \\
\mbox{with } \bm h_L(D) & = &  \left[  \frac{h_0(kD)}{h_0'(kr_L)},\, \frac{h_1(kD)}{h_1'(kr_L)},\,  \frac{h_1(kD)}{h_1'(kr_L)},\, \frac{h_1(kD)}{h_1'(kr_L)},...,\,  \frac{h_{N_L}(kD)}{h_{N_L}'(kr_L)} \right], \nonumber
\end{eqnarray}
and
\begin{eqnarray}\label{eq:MIMO1t}
\bm Q_L & = & \mbox{diag}[\bm q_L],\,\,  \\
\mbox{with } \bm q_L & = & [q(0,\cos\alpha),\, q(1,\cos\alpha),\, q(1,\cos\alpha),\, q(1,\cos\alpha),...,\, q(N_L,\cos\alpha) ]. \nonumber
\end{eqnarray}
From this point, for convenience of writing, $\bm G_s(k)$ is written as $\bm G_s$.  

Next, a SH description of the system is introduced, which is simpler and more appropriate for the analysis performed in the following chapters. 
For this purpose, two additional vectors are defined;
$\bm p_{SH}$, the SH coefficients of the pressure around the microphone array with dimensions of $(N_M+1)^2\times 1$, and $\bm u_{SH}$, the SH coefficients of the continuous normal velocity of the loudspeaker array with dimensions of $(N_L+1)^2\times 1$, given by: 
\begin{eqnarray}\label{eq:MIMO3}
\bm p & = & \bm Y_M \bm p_{SH}  \\ 
\bm u_{SH} & = & \frac{4\pi}{L} \bm Y_L \bm u. \label{eq:MIMO4}
\end{eqnarray} 
The system represented in SH, $\bm p_{SH} = \bm G \bm u_{SH}$, is formulated by multiplying $\bm G_s$ with the inverse discrete SFT of the microphone array, $\frac{4\pi}{M}\bm Y_M^\mathrm{H}$, from the left, and by multiplying it with the pseudo-inverse solution of eq.~\eqref{eq:MIMO4}, $\bm Y_L$, from the right, yielding: 
\begin{eqnarray}\label{eq:sysPWDM2}
\bm G & = &  \bm B_M \bm y^{*}(\bm \eta_{ML})\bm y^\mathrm{H}(\bm \theta_{LM}) \bm H_L(D) \bm Q_L. 
\end{eqnarray}  

This compact system representation can be interpreted in the following way: 
$\bm y(\bm \theta_{LM})$ and $\bm y(\bm \eta_{ML})$ represent the relative directions of the individual loudspeakers and microphones in the respective reference frames, $\bm H_L(D)$ represents the propagation from the loudspeaker array to the microphone array, $\bm Q_L$ expresses the effects of the spherical cap model employed at the loudspeaker array \cite{rafaely2009spherical}, and $\bm B_M$ represents the propagation matrix of the microphone array. 
From this point on in the paper only the SH representation of the system is used, as in eq.~\eqref{eq:sysPWDM2}.

\section{FAR-FIELD PROPERTIES OF THE MIMO SYSTEM IN FREE FIELD}\label{sec:sys_properties_freefield}

Two properties of the system in free-field conditions are studied in this section. 
These properties will be useful for formulating and analyzing a MIMO system model for enclosures in the next section. 

\subsection{System rank and singular vectors}\label{sec:sys_rank_freefield}

The system's matrix representation in eq.~\eqref{eq:sysPWDM2} shows its inherent unit rank; 
the matrix is formulated as an outer product of two vectors, $\bm B_M \bm y^{*}(\bm \eta_{ML})$ and $\bm Q_L^\mathrm{H} \bm H_L^\mathrm{H} \bm y^\mathrm{*}(\bm \theta_{LM})$, 
which are proportional to the matrix left and right singular vectors, respectively, due to the unit rank of the matrix.  
Therefore, the system matrix describing the transfer function between the loudspeaker array and the microphone array holds information regarding the directions of arrival and radiation. 
Moreover, the rank of the matrix stays unity regardless of the microphone array configuration, i.e. regardless of $\bm B_M$. 
This implies that a scattered field, present when using rigid microphone arrays, does not change the rank of the system, for it is determined entirely by the incident field. 
Thus, it is enough to regard only the incident field when studying the behavior of a system. 

\subsection{Effects of rotation and mirroring}\label{subsec:ROT}

The effects of rotation and mirroring are important when considering a system model for rooms, as formulated by the image method, for example.   
In this case, the system can be represented as a sum of free-field systems, which incorporate mirrored and rotated images of the loudspeaker array. 

Given a function on a sphere, $f$, represented in SH as in eq.~\eqref{eq:back_SFT}, rotation is achieved by applying Wigner-D rotation functions \cite{varshalovich1987quantum}. 
The rotated function, $f^\mathrm{R}(\bm \theta)$, is given by modifying the SH coefficients: 
\begin{eqnarray}\label{eq:DWIGNER0}
f^\mathrm{R}(\bm \theta) & = & \sum_{n=0}^{\infty}\sum_{m = -n}^{n}f_{nm}^\mathrm{R} Y_n^m(\bm \theta), \; \;\;\;
 f_{nm}^\mathrm{R}  =   \sum_{m'=-n}^{n} f_{nm'} D_{mm'}^{n}(\alpha, \beta, \gamma),
\end{eqnarray}
in which $D_{mm'}^{n}$ is the Wigner-D function and $(\alpha, \beta, \gamma)$ are the Euler rotation angles\cite{varshalovich1987quantum}. 

For order-limited functions, Wigner-D rotation matrices can be constructed by assembling Wigner-D functions for all $n$ and $m$;
i.e., for a function given by its coefficients vector with a SH order of $N$, $\bm f_{SH}$, the rotated vector of coefficients, $\bm f_{SH}^\mathrm{R}$ is given by: 
\begin{eqnarray}\label{eq:DWIGNERvector}
\bm f_{SH}^\mathrm{R} & = & \bm D_N(\alpha, \beta, \gamma)  \bm f_{SH},
\end{eqnarray}  
where $\bm D_N$, short for $\bm D_N(\alpha, \beta, \gamma)$, is a block-diagonal unitary matrix with dimensions of $(N+1)^2\times(N+1)^2$, see ref. [30]. 

Rotating a spherical loudspeaker array or a spherical microphone array is performed in the SH domain and is given by multiplying $\bm G$ with a rotation matrix on the right or on the left, respectively. 
For example, for a system with a rotated loudspeaker array, the new transfer matrix, $\bm G^{\mathrm{R}}$, is given by: 
\begin{eqnarray}\label{eq:DWIGNER1}
\bm G^{\mathrm{R}} &=& \bm G \bm D_{N_L}^\mathrm{H}. 
\end{eqnarray}
Similarly, for a system with a rotated microphone array, $\bm G$ is multiplied from the left with a rotation matrix with suitable dimensions. 

\begin{figure}[t]
\centering
\includegraphics[width=0.4\linewidth]{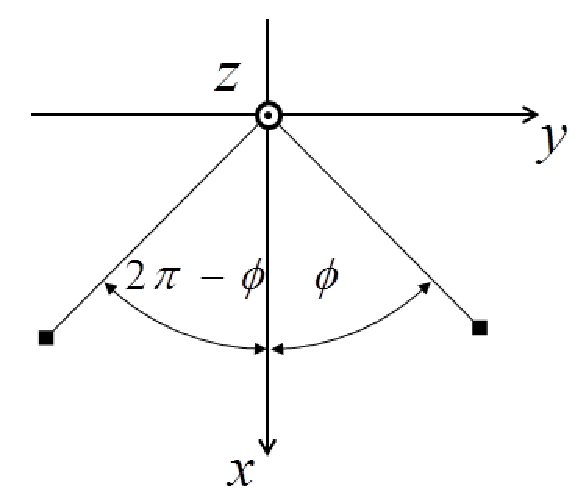}
\caption[Mirroring about the x-z plane, viewed from the z-axis (top view)]{Mirroring about the x-z plane, viewed from the z-axis (top view)}
\label{fig:mirroringyaxis}
\end{figure}

Mirroring operators are used in current simulation software \cite{wabnitz2010room,ajaj2008software,schroder2010virtual} and in multi-channel systems such as Ambisonics\cite{svensson2004use}. 
Mirroring is easily described using a Cartesian coordinate systems by inversing one of the axes, resulting in a unitary operator; i.e., for a vector given in Cartesian coordinates, $[x, y, z]^\mathrm{T}$, mirroring about the x-z plane is applied by multiplying with the unitary matrix, $\mbox{diag}(1,-1,1)$. 
For polar coordinate systems, mirroring about the x-z plane is implemented by inserting $(2\pi-\phi)$, instead of $\phi$, as the azimuth argument in a spherical harmonic function. 
This is demonstrated in fig.~\ref{fig:mirroringyaxis}. 
Due to the symmetry of spherical harmonic functions, mirroring about the x-z plane results in \cite{driscoll1994computing}: 
\begin{eqnarray}\label{eq:MIRRORING}
Y_n^m\left(\theta, 2\pi - \phi \right)  =   Y_n^{m^*}\left(\theta,  \phi \right) = (-1)^m Y_n^{-m}\left(\theta,  \phi \right).
\end{eqnarray}
Now, mirroring of a function can be formulated by rearranging the function's spherical SH coefficients: 
\begin{eqnarray}\label{eq:MIRRORINGa}
f^\mathrm{M}(\theta,\phi) & = & f(\theta,2\pi-\phi) = \sum_{n=0}^{\infty}\sum_{m = -n}^{n}f_{nm} Y_n^m(\theta,2\pi-\phi) \nonumber \\ 
& = & \sum_{n=0}^{\infty}\sum_{m = -n}^{n}f_{nm} (-1)^m Y_n^{-m}(\theta,\phi), 
\end{eqnarray}
and by changing the order of the summation over $m$ and defining $f_{nm}^\mathrm{M} =  (-1)^m f_{n(-m)}$, eq.~\eqref{eq:MIRRORINGa} can be written as: 
\begin{eqnarray}\label{eq:MIRRORINGb}
f^\mathrm{M}(\theta,\phi) & = & \sum_{n=0}^{\infty}\sum_{m = -n}^{n}f_{nm}^\mathrm{M} Y_n^m(\theta,\phi).
\end{eqnarray}

For order-limited functions, mirroring matrices can by constructed following the definition of $f_{nm}^\mathrm{M}$, and for a matrix representation of a vector with a SH order of $N$, $\bm f_{SH}$, the mirrored vector, $\bm f^\mathrm{M}_{SH}$, is given by: 
\begin{eqnarray}\label{eq:MIRRORINGvector}
\bm f_{SH}^\mathrm{M} & = & \bm M_N \bm f_{SH},
\end{eqnarray}  
where $\bm M_N$ is a block-diagonal unitary matrix with dimensions of $(N+1)^2\times (N+1)^2$. 
$\bm M_N$ is composed of two matrices:
\begin{eqnarray}
\label{eq:MIRRORING4a}
\bm M_N &=& \bm M_{-1}(N) \bm M_{\mathrm{perm}}(N),
\end{eqnarray}
in which $\bm M_{-1}(N) = \mbox{diag}[1, -1, ... ,(-1)^N]$ is a diagonal matrix and $\bm M_{\mathrm{perm}}(N)$ is a block-diagonal matrix with $(N+1)$ blocks, in which every block of dimensions of $(2n+1)$, for $n=0,1,...,N$, is an anti-diagonal unit matrix. 

Mirroring a spherical loudspeaker array or a spherical microphone array is performed in the SH domain by multiplying $\bm G$ with a mirroring matrix on the right or on the left, respectively, similar to the rotation operators from eq.~\eqref{eq:DWIGNER1}. 
Mirroring about other planes can be implemented by concatenating rotation and mirroring matrices; 
for example, using eqs.~\eqref{eq:DWIGNERvector} and \eqref{eq:MIRRORINGvector}, an order-limited function mirrored about the y-z plane is denoted $\tilde{\bm f}_{SH}$ and calculated by: 
\begin{eqnarray}\label{eq:MIRRORING6}
\tilde{\bm f}_{SH} & = &  \bm D_N(-\frac{\pi}{2}, 0, 0)\bm M_N \bm D_N(\frac{\pi}{2}, 0, 0) \bm f_{SH}, 
\end{eqnarray} 
in which $D_N(\frac{\pi}{2}, 0, 0)$ applies counter-clockwise rotation by $\frac{\pi}{2}$ about the z-axis, $\bm M_N$ mirrors about the x-z plane, and $\bm D_N(-\frac{\pi}{2}, 0, 0)$ applies clockwise rotation by $\frac{\pi}{2}$ about the z-axis. 
Following this, mirroring about any axes is a unitary operator, for it can be implemented as a sequence of unitary operators. 

Since rotation and mirroring matrices are unitary, multiplying $\bm G$ with these matrices on either side does not affect the nature of the singular values of $\bm G$.  
In terms of control theory, this implies that for rotation and mirroring of either arrays, only the loudspeaker and microphone ``directions", the singular vectors, are rotated, and the energetic properties of the system, represented by the singular values, remain the same. 

\section{MIMO SYSTEM MODEL IN ROOMS}\label{sec:FF2RTF}

An acoustic MIMO system within a rectangular room can be described as a sum of free-field MIMO systems, following Allen and Berkley's image method\cite{allen1979image}.
The RIR is described by replacing the walls with reflected sources, referred to as image sources. 
Originally proposed for omnidirectional sources, the image method can be generalized for spherical loudspeaker and microphone arrays with the main difference being that image sources are not only displaced, but also mirrored according to their reflection paths.

The new system is denoted $\bm G^{\texttt{Room}}$ and is given by summation over all image sources: 
\begin{eqnarray}\label{eq:sysFFMIMORTF}
\bm G^{\texttt{Room}} & = & \sum_{g} A_g \bm B_M  \bm y_{M,g}^{*}\bm y_{L,g}^\mathrm{H} \bm H_L(D_g) \bm Q_L ,
\end{eqnarray}
where $g$ is the index of the reflection/image source, $A_g$ is a complex constant expressing the attenuation and delay of the $g^{th}$ reflection path, $\bm y_{L,g}$  and $\bm y_{M,g}$ are the steering vectors of the loudspeakers and microphone arrays, as in eq.~\eqref{eq:sysPWDM1}, with the propagation and arrival directions of the $g^{th}$ reflection, $\bm \theta_{LM,g}$ and $\bm \eta_{ML,g}$, respectively, and $\bm H_L(D_g)$ is the propagation matrix, as in eq.~\eqref{eq:sysFFMIMO2}, with the distance between the microphone array and the $g^{th}$ image source, $D_g$, as the argument. 
As can be seen, all free-field systems representing the different images are described using the simplified far-field model. 
This assumption is fair, for if the distance between the microphone and loudspeaker arrays is large enough, a reasonable assumption for real rooms, then the distances between the microphone array and all image sources are even larger.  

The rank of a MIMO system in a room is bounded by the number of significant reflections, and the subspace spanned by the left and right singular vectors (separately) is equivalent to the subspace spanned by the steering vectors of the microphone and loudspeaker arrays, respectively. 
Following the analysis of free-field systems, each element from the summation in eq.~\eqref{eq:sysFFMIMORTF} has unit rank, and the summation over unit-rank matrices potentially increases the rank of the matrix. 
In particular, the rank of the matrix is bounded by the minimum of the number of reflections and the dimensions of the system, i.e. $min(I,(N_M+1)^2,(N_L+1)^2 )$, where $I$ is the number of significant reflections.
Therefore, in a real room with a high number of significant reflections, the rank is expected to be high, facilitating greater control of the sound field around the microphone array using the loudspeaker array, which can be exploited for improved sound field analysis\cite{morgenstern2013enhanced}. 

\section{SIMULATION STUDY}\label{sec:SimStud}

A simulation study is performed with the aim of validating the theoretical results regarding the rank and the spatial diversity of an acoustic MIMO system in a room.
Spherical loudspeaker and microphone arrays with SH orders of $N_L= 5$ and $N_M= 5$ were simulated, located at $(x_L, y_L,z_L) = (1,1.5,2)$m and $(x_M, y_M,z_M) = (7,6.5,6)$m, respectively, within a room with dimensions of $(8,9,10)$m, which was simulated using the McRoomSim software\cite{wabnitz2010room}.   
Multiple transfer functions were simulated, relating the SH coefficients of the microphone array to the SH coefficients of the loudspeaker array, and arranged in matrix form, as in eq.~\eqref{eq:sysFFMIMORTF}, and time-domain RIR's were calculated using the inverse discrete Fourier transform (FT). 
Several rooms were simulated with the same dimensions, differing only in the walls' reflection coefficients.
Frequency-dependant reflection coefficients were simulated within the range of $[0.005, 0.6]$. 

The (mathematical) rank of a matrix is an integer quantity by definition that does not take into account the full singular value spectrum, and can thus be very sensitive to noise. 
The effective rank of a matrix\cite{roy2007effective}, on the other hand, takes into account the distribution of the singular values, and is thus more suitable for measuring the effective dimensionality of real systems that encounter noise. 
For a matrix, $\bm A$, with dimensions of $M\times N$, the effective rank is defined as:
\begin{eqnarray}\label{eq:EFFECTIVE RANK}
\text{erank}(A) & \triangleq & e^{\mathit{W}(\sigma_1,\sigma_2,..,\sigma_Q)}, 
\end{eqnarray} 
where $Q = min(N,M)$, $\sigma_i$ are normalized singular values, defined as: 
\begin{eqnarray}\label{eq:EFFECTIVE RANK2}
\sigma_i & = & \frac{s_i}{\sum_{l=1}^{Q}s_l}, 
\end{eqnarray} 
$ \{s_i\}_{i =1}^{Q}$ are the non-negative singular values and $\mathit{W}(\sigma_1,\sigma_2,..,\sigma_Q)$ is the entropy, given by: 
\begin{eqnarray}\label{eq:EFFECTIVE RANK3}
\mathit{W}(\sigma_1,\sigma_2,..,\sigma_Q) & = & -\sum_{l=1}^{Q} \sigma_l \log{\sigma_l}.  
\end{eqnarray} 
For more properties of the effective rank, the reader is referred to ref. [35]. 

The effective rank of the system is calculated for an operating frequency of $f=700$Hz and for time segments with an increasing duration of the RIR in the following manner:
\begin{enumerate}
\item First, discrete-time RIRs are calculated using an inverse FT applied on the elements of the system transfer function $\bm G(k)$ from eq.~\eqref{eq:sysclassic}, using a sampling frequency of $48$kHz.
These are assembled into tensor form, $\bm G[t]$, with dimensions of $(N_M+1)^2\times(N_L+1)^2\times(T)$, where $t$ is the discrete-time index and $T$ is the length of the discrete-time RIR. 
\item Then, the elements of $\bm G[t]$ are multiplied by windows that start at $t=0$ and have an increasing duration. 
This results in tensors denoted $\bm G[\tau, t]$, where $\tau$ is the duration of the time-window. 
$\bm G[\tau, t]$ are calculated for values of $\tau$ starting at $\tau=0$ and ending with $\tau=T$. 
Note that for $\tau = T$ the whole response is included, i.e. $\bm G[T, t] = \bm G[t]$.  
\item Finally, a discrete FT is applied on the elements of $\bm G[\tau,t]$ for the operating frequency $f=700$Hz, resulting in matrices $\bm G[\tau, f]$ with dimensions of $(N_M+1)^2\times(N_L+1)^2$, calculated for the different values of $\tau$. 
Note that for $\tau= T$ this results in a regular discrete FT, i.e. $\bm G[T, f] = \bm G(\frac{2\pi f}{c})$.  
\end{enumerate}

\begin{figure}[t]
\centering
\begin{subfigure}{.55\textwidth}
  \centering
  \includegraphics[width=1\linewidth]{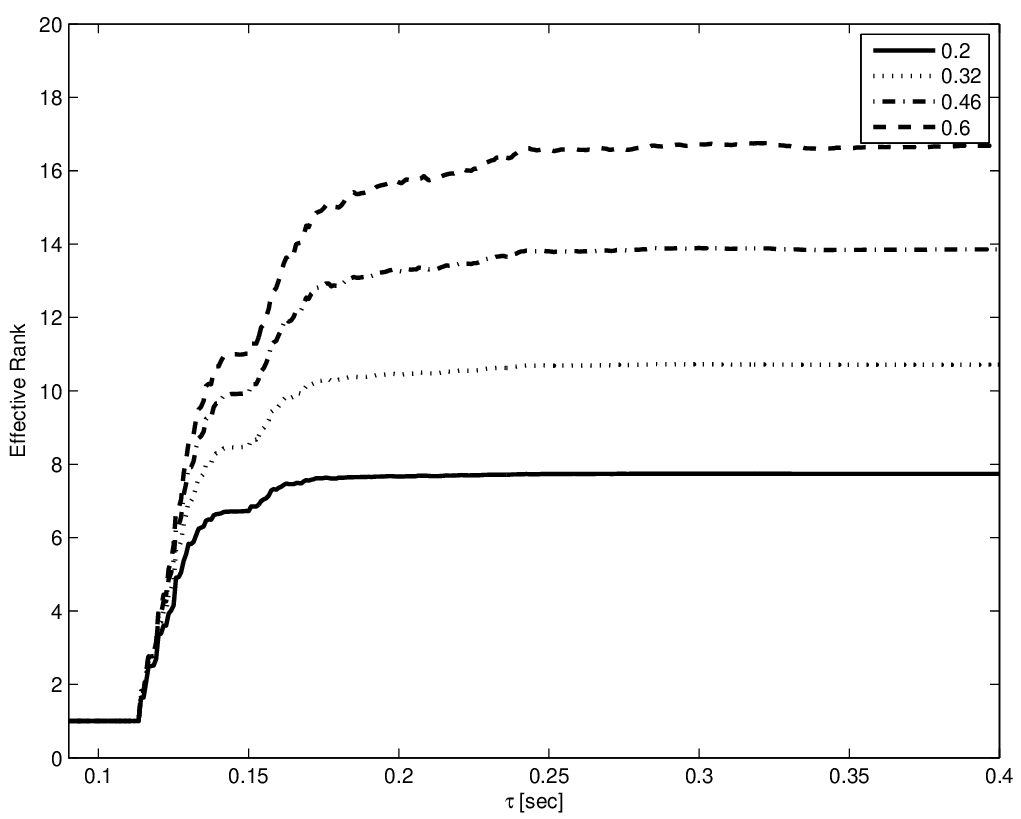}
  \caption{} 
\end{subfigure}%
\\
\begin{subfigure}{.55\textwidth}
  \centering
  \includegraphics[width=0.9\linewidth]{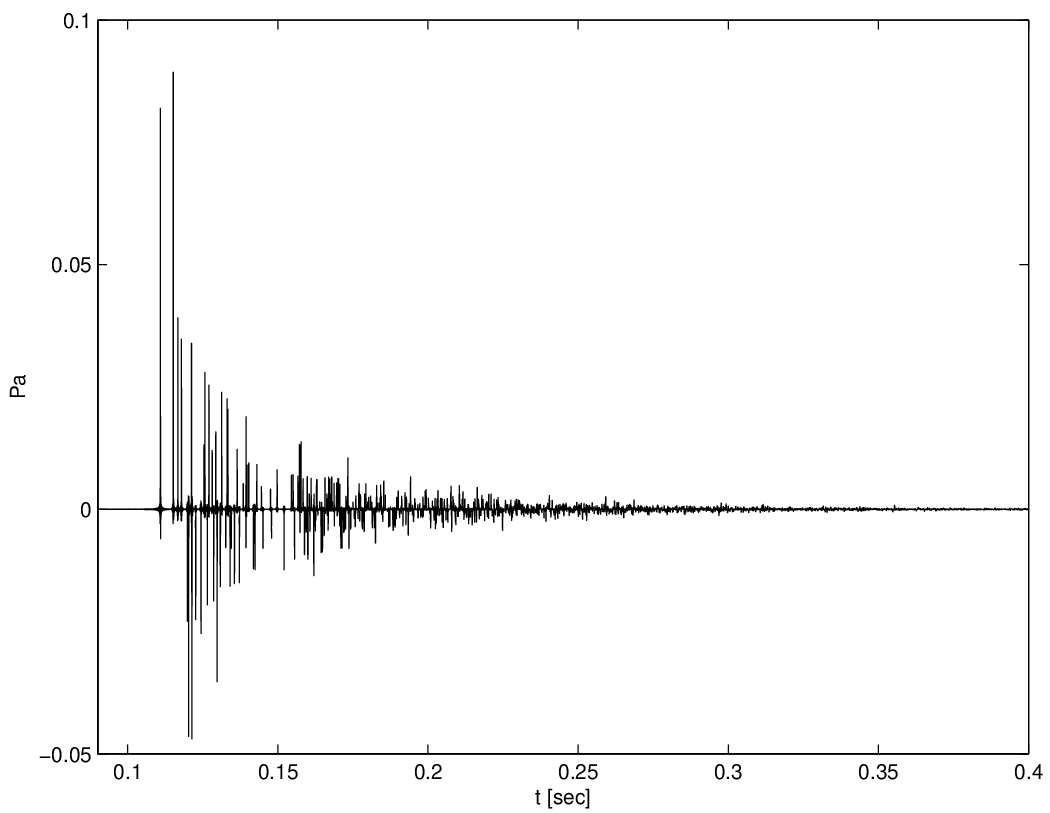}
  \caption{} 
\end{subfigure}
\caption{(color online) (a) System effective rank at $700$Hz as a function of $\tau$, time-window duration, for a room with adjustable reflection coefficients, as detailed in the figure caption. (b) Simulated impulse response between omnidirectional loudspeaker and microphone.}
\label{fig:sim_effectiverank}
\end{figure}

Fig.~\ref{fig:sim_effectiverank} presents the effective rank of $\bm G[\tau, f]$ as a function of $\tau$, the time-window length, calculated for rooms with different wall reflection coefficients. 
The omnidirectional RIR is given as a reference. 
The effective rank is shown to start at unity for small values of $\tau$ that include only the direct sound, and to increase as $\tau$ increases, resulting in the accumulation of more reflections.  
Moreover, it is evident that rooms with higher reflection coefficients reach higher effective ranks in the steady-state, i.e. for $\tau = T$.  

To demonstrate the implications of a system's effective rank, the ability of shaping the sound field around the microphone array by controlling the loudspeaker array directivity is simulated and tested as a function of the effective rank. 
A target incident field around the microphone array is chosen, denoted as $\bm p^\mathrm{t}$, 
and SH coefficients of the loudspeaker array that achieve the target pattern are sought. 
Mathematically, loudspeaker SH coefficients that give an exact solution are available if $\bm G(\frac{2\pi f}{c})$ is invertible, but practically, this may yield an ill-conditioned solution. 
Thus, an inverse system is found using the singular value decomposition by inverting only non-zero singular values.
For a robust solution, only singular values that are smaller by 50dB or less than the largest singular value are inverted, i.e. singular values higher than $s_1\times 10^{-\frac{50}{20}}$,  where $s_1$ is the first singular value of the system. 
The inverse of the other singular values is set to zero. 
The robust inverse system is denoted $\bm G^\dagger$, and the loudspeaker SH coefficients, $\bm u^{\mathrm{t}}$, are given using:
\begin{eqnarray}\label{eq:simstudy1}
\bm u^{\mathrm{t}}& = &  \bm {G}^{\dagger} \bm p^\mathrm{t}. 
\end{eqnarray} 
The resulting reproduced sound field, $\bm p^{\mathrm{a}}$, is calculated by substituting $\bm u^{\mathrm{t}}$ in eq.~\eqref{eq:sysclassic}, by:  
\begin{eqnarray}\label{eq:simstudy2}
\bm p^{\mathrm{a}} & = &  \bm G \bm u^{\mathrm{t}} \nonumber \\
& = &  \bm G \bm {G}^{\dagger} \bm p^\mathrm{t}. 
\end{eqnarray}

A normalized reproduction error can be defined by: 
\begin{eqnarray}\label{eq:simstudy3}
E & = & \frac{||\bm p^{\mathrm{a}}- \bm p^{\mathrm{t}}||}{||\bm p^{\mathrm{t}}||}, 
\end{eqnarray}
where $||\cdot||$ is the Euclidean norm. 

\begin{figure}[t]
\centering
\begin{subfigure}{.5\textwidth}
  \centering
  \includegraphics[width=1\linewidth]{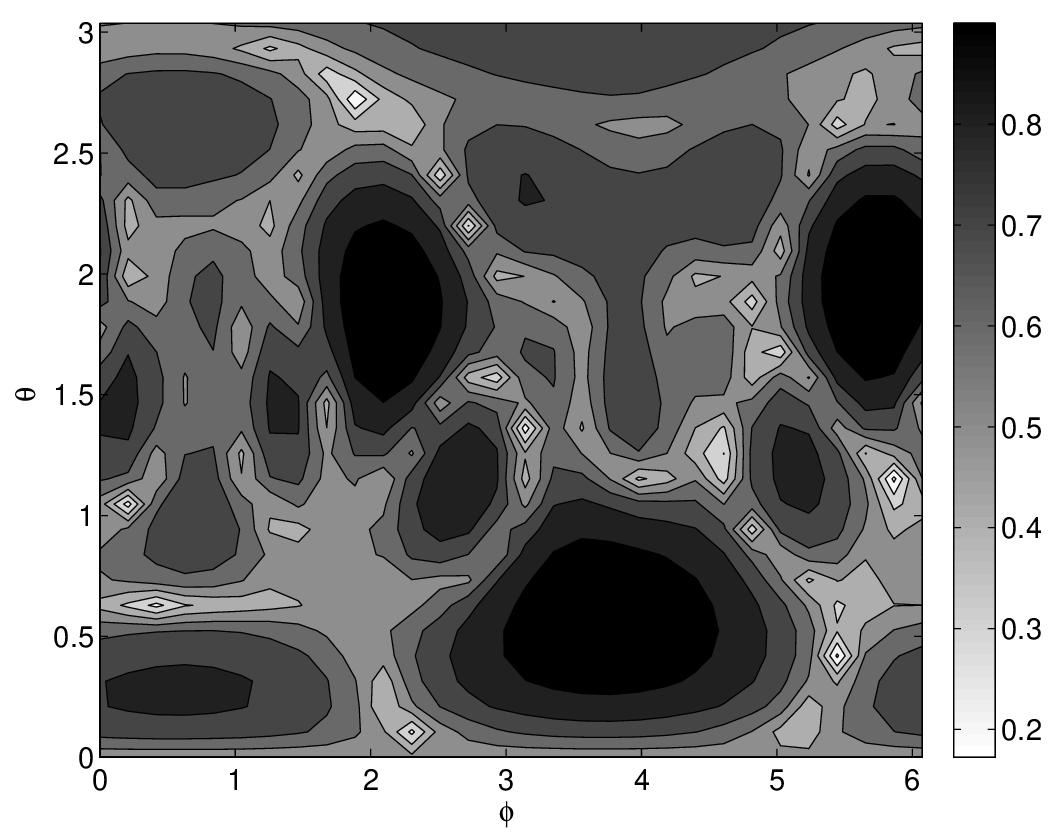}
  \caption{} 
\end{subfigure}%
\begin{subfigure}{.5\textwidth}
  \centering
  \includegraphics[width=1\linewidth]{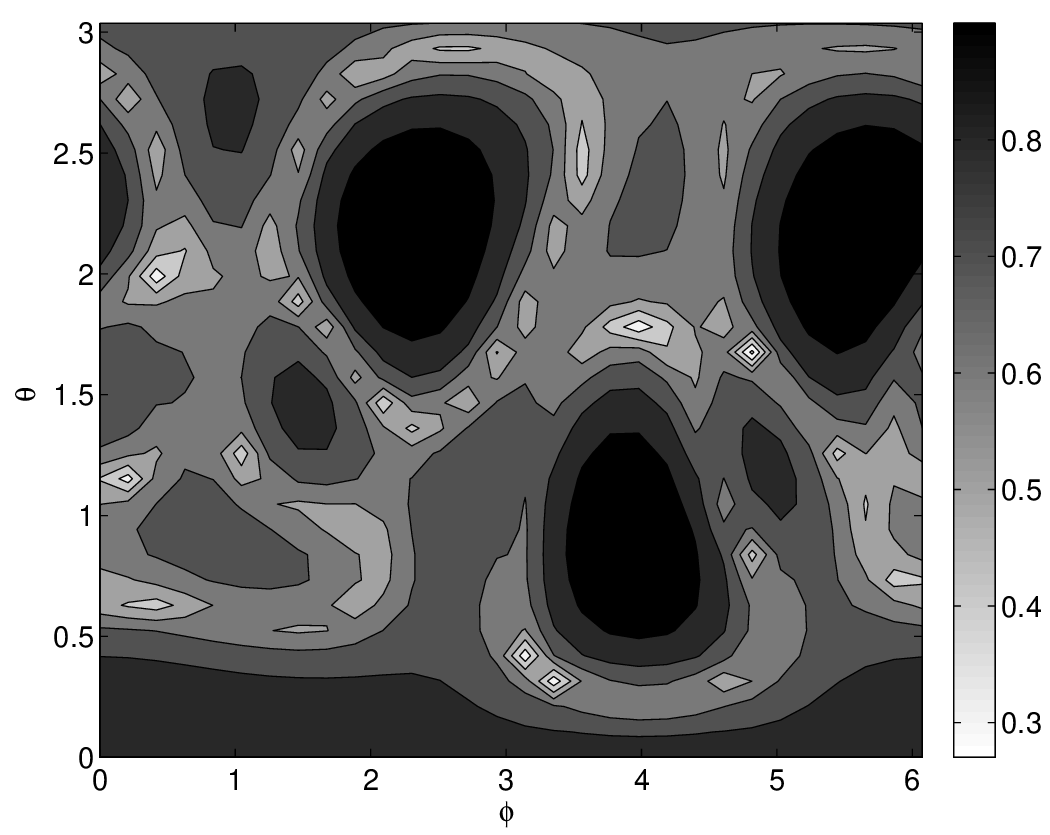}
  \caption{} 
\end{subfigure}
\begin{subfigure}{.5\textwidth}
  \centering
  \includegraphics[width=1\linewidth]{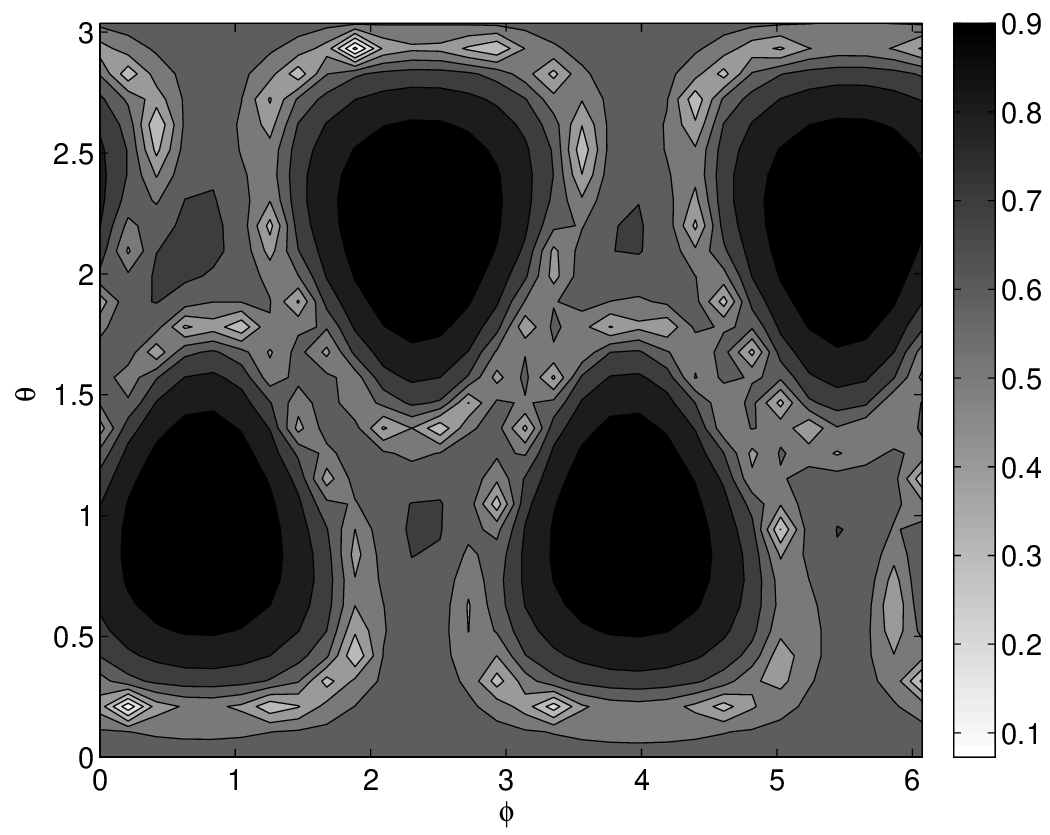}
  \caption{} 
\end{subfigure}%
\begin{subfigure}{.5\textwidth}
  \centering
  \includegraphics[width=1\linewidth]{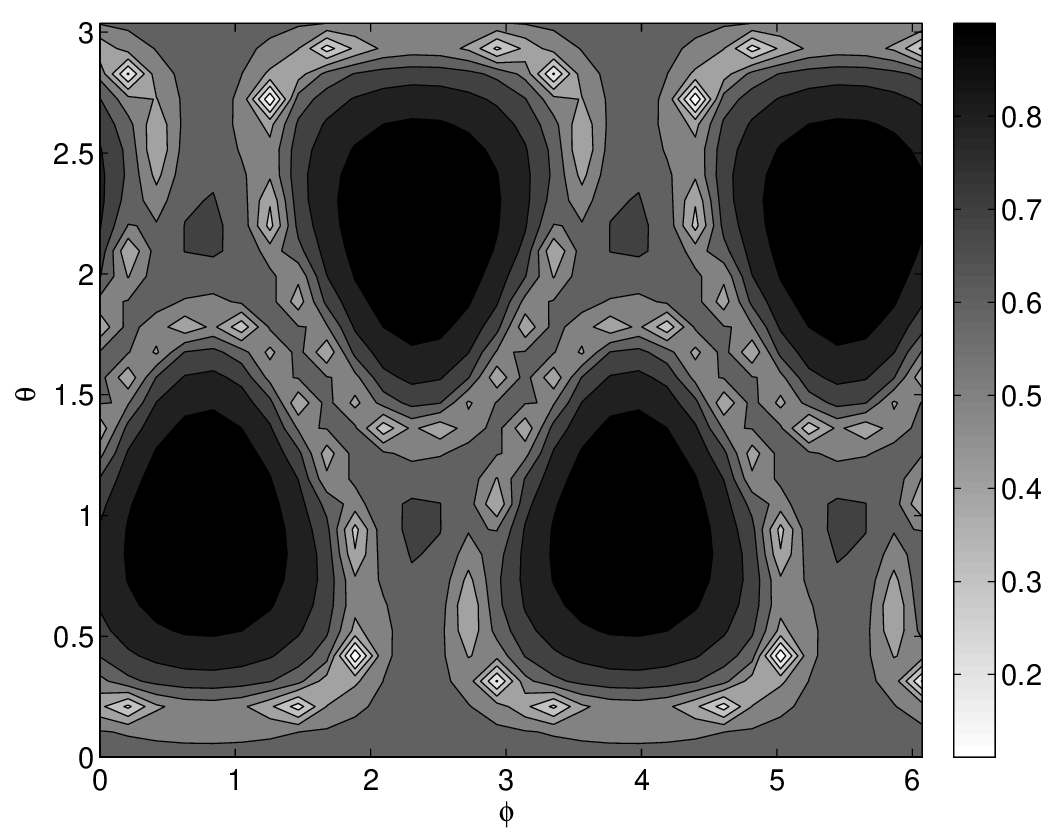}
  \caption{} 
\end{subfigure}
\caption[Reproduced sound field around the microphone array: (a)-(c) Simulated rooms, see table.~\ref{table:reproduction} (d)- Target incident field.]{Reproduced sound field around the microphone array: (a)-(c) Simulated rooms, see table 
 1. (d)- Target incident field.}
\label{fig:simulation_effective_rank2}
\end{figure}

The possibility of shaping the sound field around the microphone array is studied by first setting a target incident field with high spatial complexity; 
a target incident field comprising 12 orthonormal vectors was chosen to avoid trivial incident fields which can be achieved using systems with low effective rank.  
The loudspeaker array SH coefficients were calculated for different reflection coefficients, following eq.~\eqref{eq:simstudy1}. 
Fig.~\ref{fig:simulation_effective_rank2} presents the target field and the achieved fields, and the corresponding normalized reproduction error and reflection coefficients are presented in table 
1. 
It is evident that as the value of the reflection coefficients is increased, a more accurate reproduction of the target field is obtained.  
This implies that significant reflections contribute to the achievable spatial complexity of the incident field around a microphone array in a room.

\section{EXPERIMENTAL INVESTIGATION}\label{sec:EXPInvestigation}

A MIMO system comprising spherical arrays was implemented and investigated experimentally. 
A measurement system was positioned in a lecture hall at Ben-Gurion University. 
A $15$cm radius measurement loudspeaker array developed at the Institute of Technical Acoustics, Aachen University, with 12 loudspeaker units was placed at the center of the stage, and a $4.2$cm radius mh acoustics' Eigenmike\textsuperscript{\textregistered} microphone array with 32 microphones was placed in the 4$^{th}$ row of the seating area. 
The auditorium has 170 seats and an approximate volume of $822m^3$; a reverberation time of $1.72$ seconds was measured by the system\cite{standard19973382}. 

The experiment was performed in the following stages: 
\begin{enumerate}
\item Each loudspeaker unit played a 5 second-long linear swept-sine signal, scanning frequencies from 0 to 5 kHz, and this was simultaneously recorded by the 32 microphones. 
\item All $12\times 32$ responses were arranged in matrix form, and were deconvolved with the swept-sine signal played back by the loudspeaker units, resulting in the room impulse responses, $\bm G[t]$. 
\end{enumerate}

\begin{figure}[t]
\centering
\begin{subfigure}{.50\textwidth}
  \centering
  \includegraphics[width=1\linewidth]{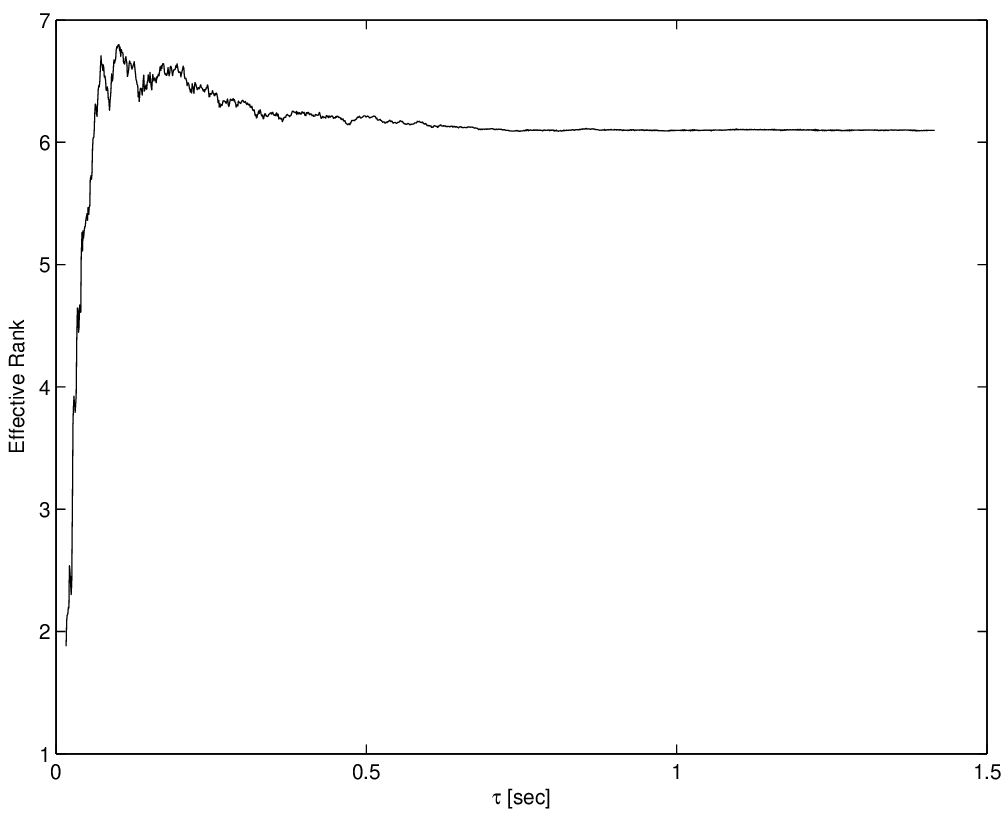}
  \caption{} 
\end{subfigure}%
\\
\begin{subfigure}{.50\textwidth}
  \centering
  \includegraphics[width=1\linewidth]{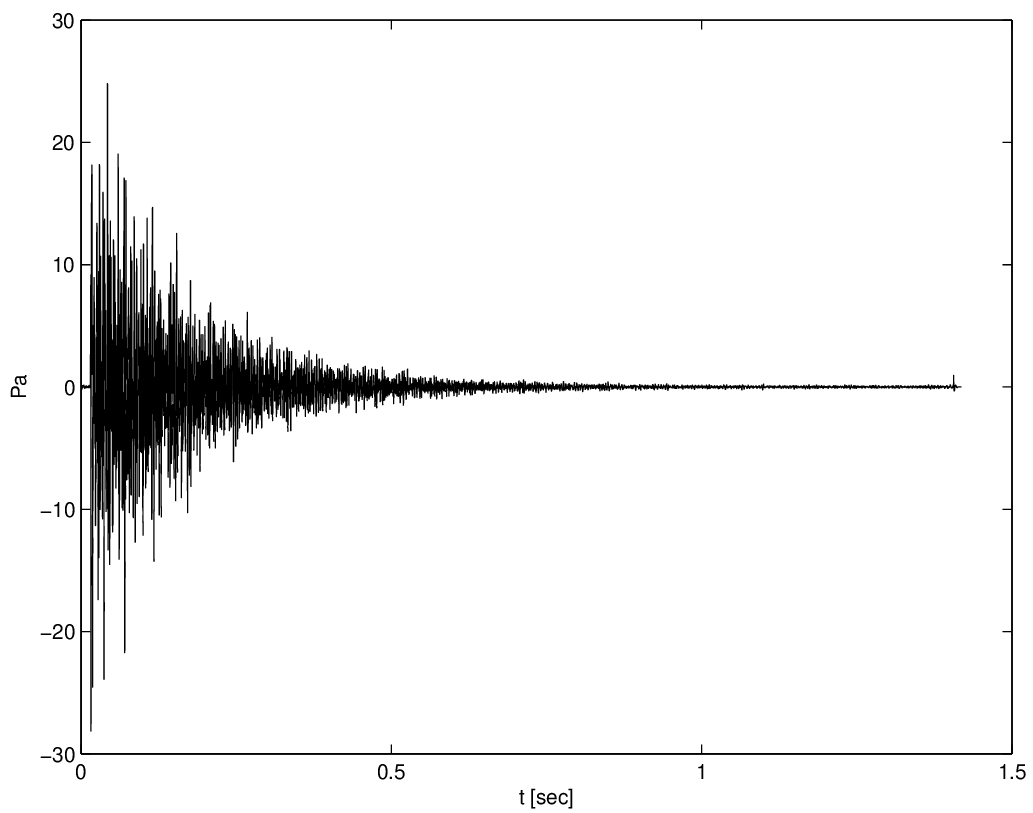}
  \caption{} 
\end{subfigure}
\caption[(color online) (a) System effective rank as a function of time-window length $\tau$. (b) Measured impulse response between omnidirectional directivity patterns at loudspeaker and microphone arrays.]{(color online) (a) System effective rank as a function of time-window length $\tau$. (b) Measured impulse response between omnidirectional directivity patterns at loudspeaker and microphone arrays.}
\label{fig:exp_effectiverank}
\end{figure}

The effective rank of the measured system was calculated for an operating frequency of $f=700$Hz and for increasing time segments of the RIR, as in the case of the simulation study, and is denoted $\bm G[\tau, f]$. 
Fig.~\ref{fig:exp_effectiverank} presents the system effective rank as a function of the time-window length, $\tau$, and the omnidirectional RIR is given as a reference. 
As in the simulation, the effective rank is shown to start close to unity for the direct sound and increases as the reflections in the room become denser. 
Additionally, it seems that the effective rank is bounded by a value smaller than 7; 
this does not mean that there are a maximum of 7 reflections at all time bins, just that the effective SH orders of the loudspeaker and microphone arrays, which are a function of the operating frequency, bound the efficient rank in this case. 
Unlike the simulation, the effective rank of the measured system is shown to decrease at some point in the diffuse segment of the response. 
This implies that at some point, adding more and more reflections changes the singular value distribution resulting in a decrease in the system effective rank.   	

\begin{figure}[t]
\centering
\includegraphics[width=0.7\textwidth]{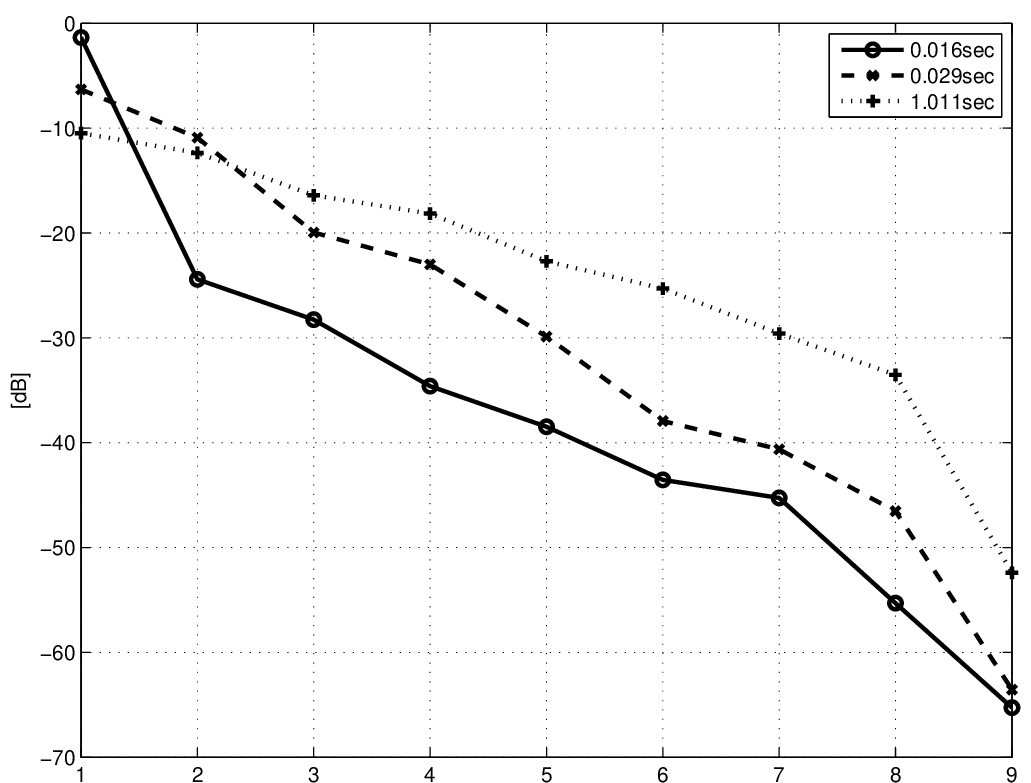}
\caption{(color online) Singular value distribution of $\bm G[0.016\text{sec}, 700\text{Hz}]$, $\bm G[0.029\text{sec}, 700\text{Hz}]$, and $\bm G[1.011\text{sec}, 700\text{Hz}]$, with effective ranks of 1.88, 3.92, and 6.09, respectively.}
\label{fig:exp_singulardistribution}
\end{figure}

The singular value distribution of $\bm G[\tau, f]$ is calculated and presented in fig.~\ref{fig:exp_singulardistribution} for three different time-window durations, $\tau = 0.016$sec, which includes only the direct sound, $\tau = 0.029$sec, which includes additional early reflections, and $\tau = 1.011$sec, which includes most of the response. 
It can be seen that for the short time-window that includes only the direct sound, there is a significant first singular value, higher than the first singular values found in longer duration time-windows. 
On the other hand, the remaining singular values for that time window are smaller than the singular values of longer duration time-windows. 
This demonstrates the implications of the effective rank; for long time-windows the effective rank increases, resulting in a more uniform distribution of the singular values.     
 
\section{CONCLUSIONS}\label{sec:CONC}

In this paper, a theoretical formulation and analysis of an acoustic MIMO system comprising a spherical loudspeaker array and a spherical microphone array was presented. 
System models were developed and studied for free-field conditions and for rooms.
The analysis shows that singular values remain unchanged under rotation and mirroring of the arrays. 
This is important when considering positioning of arrays in rooms. 
Moreover, the effective rank of a system was shown to increase as more reflections are taken into account in the RIR. 
This suggests that in rooms, a sound field with high spatial complexity can be shaped around a microphone array, modeling a human listener, by controlling the weights of the loudspeaker array. 
It also suggests that the use of advanced signal processing methods, which require high rank matrices, may be possible using the proposed MIMO system. 
Future work includes the design of new methods for spatial analysis and for sound field synthesis using the proposed MIMO system.  

\begin{thebibliography}{10}

\bibitem{standard19973382}
ISO, ``3382. acoustics--measurement of the reverberation time of rooms
  with reference to other acoustical parameters,'' {International Standards
  Organization} (1997).

\bibitem{marshall2001spatial}
A.~Marshall and M.~Barron, ``Spatial responsiveness in concert halls and the
  origins of spatial impression,'' { Applied acoustics}, vol.~62, no.~2,
  pp.~91--108 (2001).

\bibitem{griesinger1998general}
D.~Griesinger, ``General overview of spatial impression, envelopment,
  localization, and externalization,'' in {Audio Engineering Society
  Conference: 15th International Conference: Audio, Acoustics \& Small Spaces},
  Audio Engineering Society, Copenhagen, Denmark, 1998.

\bibitem{morimoto2001role}
M.~Morimoto, K.~Iida, and K.~Sakagami, ``The role of reflections from behind
  the listener in spatial impression,'' {Applied Acoustics}, vol.~62,
  no.~2, pp.~109--124 (2001).

\bibitem{gover2002microphone}
B.~N. Gover, J.~G. Ryan, and M.~R. Stinson, ``Microphone array measurement
  system for analysis of directional and spatial variations of sound fields,''
  {The Journal of the Acoustical Society of America}, vol.~112, no.~5,
  pp.~1980--1991 (2002).

\bibitem{rafaely2005analysis}
B.~Rafaely, ``Analysis and design of spherical microphone arrays,'' {IEEE Transactions on Speech
  and Audio Processing}, vol.~13, no.~1, pp.~135--143 (2005).

\bibitem{balmages2004room}
I.~Balmages and B.~Rafaely, ``Room acoustics measurements by microphone
  arrays,'' in {23rd IEEE Convention of Electrical and Electronics Engineers in Israel, 2004.
  Proceedings. 2004}, pp.~420--423, IEEE.

\bibitem{tervo2009acoustic}
S.~Tervo, J.~Patynen, and T.~Lokki, ``Acoustic reflection path tracing using a
  highly directional loudspeaker,'' in {IEEE Workshop on Applications of Signal Processing
  to Audio and Acoustics, 2009. WASPAA'09.}, pp.~245--248,
  IEEE, 2009.

\bibitem{rafaely2009spherical}
B.~Rafaely, ``Spherical loudspeaker array for local active control of sound,''
  {The Journal of the Acoustical Society of America}, vol.~125, no.~5,
  pp.~3006--3017 (2009).

\bibitem{sekiguchi1992analysis}
K.~Sekiguchi, S.~Kimura, and T.~Hanyuu, ``Analysis of sound field on spatial
  information using a four-channel microphone system based on regular
  tetrahedron peak point method,'' {Applied Acoustics}, vol.~37, no.~4,
  pp.~305--323 (1992).

\bibitem{yamasaki2005measurement}
Y.~Yamasaki and T.~Itow, ``Measurement of spatial information in sound fields
  by a closely located four-point microphone method,'' {The Journal of the
  Acoustical Society of America}, vol.~84, no.~S1, pp.~S132--S132 (2005).

\bibitem{gerzon1975recording}
M.~A. Gerzon, ``Recording concert hall acoustics for posterity,'' {Journal
  of the Audio Engineering Society}, vol.~23, no.~7, pp.~569--571 (1975).

\bibitem{farina2003recording}
A.~Farina and R.~Ayalon, ``Recording concert hall acoustics for posterity,'' in
  {Audio Engineering Society Conference: 24th International Conference:
  Multichannel Audio, The New Reality}, Audio Engineering Society (2003).

\bibitem{farina2006room}
A.~Farina, ``Room impulse responses as temporal and spatial filters,'' in {The 9th Western Pacific Acoustics Conference, Seoul, Korea}, 2006.

\bibitem{morgenstern2012analysis}
H.~Morgenstern and B.~Rafaely, ``Analysis of acoustic mimo systems in enclosed
  sound fields,'' in {IEEE International Conference on Acoustics, Speech and Signal Processing (ICASSP),
  2012 }, pp.~209--212, IEEE.

\bibitem{morgenstern2012joint}
H.~Morgenstern, F.~Zotter, and B.~Rafaely, ``Joint spherical beam forming for
  directional analysis of reflections in rooms,'' {The Journal of the
  Acoustical Society of America}, vol.~131, no.~4, pp.~3207--3207 (2012).

\bibitem{morgenstern2013enhanced}
H.~Morgenstern and B.~Rafaely, ``Enhanced spatial analysis of room acoustics
  using acoustic multiple-input multiple-output (mimo) systems,'' in {Proceedings of Meetings on Acoustics}, vol.~19, 015018, Acoustical Society
  of America (2013).

\bibitem{morgenstern2014farfield}
H.~Morgenstern and B.~Rafaely, ``Far-field criterion for spherical microphone
  arrays and directional sources,'' in {Joint Workshop on Hands-free Speech Communication and
  Microphone Arrays (HSCMA), 2014}, IEEE.

\bibitem{allen1979image}
J.~B. Allen and D.~A. Berkley, ``Image method for efficiently simulating
  small-room acoustics,'' {The Journal of the Acoustical Society of
  America}, vol.~65, no.~4, pp.~943--950 (1979).

\bibitem{driscoll1994computing}
J.~R. Driscoll and D.~M. Healy, ``Computing fourier transforms and convolutions
  on the 2-sphere,'' {Advances in applied mathematics}, vol.~15, no.~2,
  pp.~202--250 (1994).

\bibitem{williams1999fourier}
E.~Williams, {Fourier acoustics: sound radiation and nearfield acoustical
  holography}, pp.~1--296
\newblock Academic Press (1999).

\bibitem{arfken1999mathematical}
G.~B. Arfken, H.~J. Weber, and D.~Spector, ``Mathematical methods for
  physicists,'' {American Journal of Physics}, vol.~67, pp.~165--169 (1999).

\bibitem{rafaely2007spatial}
B.~Rafaely, B.~Weiss, and E.~Bachmat, ``Spatial aliasing in spherical
  microphone arrays,'' {IEEE Transactions on Signal Processing}, vol.~55,
  no.~3, pp.~1003--1010 (2007).

\bibitem{meyer2000multi}
P.~S. Meyer and J.~D. Meyer, ``Multi acoustic prediction program (mapptm)
  recent results,'' presented in Reproduced Sound Conference Avon, UK, 2000. Proceedings of The Institute of Acoustics (UK), vol.~22, no.~6,
  pp.~9--16.

\bibitem{zotter2007modeling}
F.~Zotter, A.~Sontacchi, and R.~H\"{o}ldrich, ``Modeling a spherical
  loudspeaker system as multipole source,'' {Fortschritte der Akustik},
  vol.~33, no.~1, p.~221 (2007).

\bibitem{gumerov2005fast}
N.~A. Gumerov and R.~Duraiswami, {Fast multipole methods for the Helmholtz
  equation in three dimensions}, pp.~89--139.
\newblock Elsevier (2005).

\bibitem{rafaely2004plane}
B.~Rafaely, ``Plane-wave decomposition of the sound field on a sphere by
  spherical convolution,'' {The Journal of the Acoustical Society of
  America}, vol.~116, no.~4, pp.~2149--2157 (2004).

\bibitem{varshalovich1987quantum}
D.~A. Varshalovich, A.~N. Moskalev, and V.~K. Khersonskii, ``Quantum theory of
  angular momentum,'' {World Scientific},  pp.~22--25 (1987).

\bibitem{rafaely2008spherical}
B.~Rafaely and M.~Kleider, ``Spherical microphone array beam steering using
  wigner-d weighting,'' {Signal Processing Letters, IEEE}, vol.~15,
  pp.~417--420 (2008).

\bibitem{wabnitz2010room}
A.~Wabnitz, N.~Epain, C.~Jin, and A.~van Schaik, ``Room acoustics simulation
  for multichannel microphone arrays,'' in {Proceedings of the
  International Symposium on Room Acoustics, Melbourne, Australia}, 2010.

\bibitem{ajaj2008software}
R.~Ajaj, L.~Savioja, and C.~Jacquemin, ``Software platform for real-time room
  acoustic visualization,'' in {Proceedings of the 2008 ACM symposium on
  Virtual reality software and technology}, pp.~247--248, ACM, 2008.

\bibitem{schroder2010virtual}
D.~Schr{\"o}der, F.~Wefers, S.~Pelzer, D.~Rausch, M.~Vorl{\"a}nder, and
  T.~Kuhlen, ``Virtual reality system at rwth aachen university,'' in {Proceedings of the International Symposium on Room Acoustics (ISRA),
  Melbourne, Australia}, 2010.

\bibitem{svensson2004use}
U.~P. Svensson, B.~St{\o}fringsdal, A.~Solvang, and S.~Saue, ``The use of
  ambisonics in describing room impulse responses,'' in {Proceedings of the
  International Congress on Acoustics, Kyoto, Japan}, pp.~8--11, 2004.

\bibitem{roy2007effective}
O.~Roy and M.~Vetterli, ``The effective rank: A measure of effective
  dimensionality,'' {entropy}, vol.~4, p.~7 (2007).

\end{thebibliography}

\newpage
\begin{table}
\centering
\begin{tabular}[t]{|c|c|c|c|}
\hline Figure & Reflection Coefficients & $E$[dB] & Inverted Singular Values\\ 
\hline \ref{fig:simulation_effective_rank2}(a) & 0.005 & -1.58 & 7\\ 
\hline \ref{fig:simulation_effective_rank2}(b) & 0.05 & -5.014 & 20\\ 
\hline \ref{fig:simulation_effective_rank2}(c) & 0.5 & -26.29 & 35\\
\hline
\end{tabular}
\caption{Reflection coefficients, normalized errors as in eq.~\eqref{eq:simstudy3}, and number of inverted singular values.}
\label{table:reproduction}
\end{table}


\newpage

\end{document}